# Simulation of Drop Impact on a Hot Wall using SPH Method with Peng-Robinson Equation of State

*Manjil Ray, Xiufeng Yang, Song-Charng Kong*\*

*Department of Mechanical Engineering, Iowa State University, 2025 Black Engineering, Ames, IA 50011, USA*
\**Corresponding Author Email: kong@iastate.edu*

**Abstract:** This study presents a smoothed particle hydrodynamics (SPH) method with Peng-Robinson equation of state for simulating drop vaporization and drop impact on a hot surface. The conservation equations of momentum and energy and Peng-Robinson equation of state are applied to describe both the liquid and gas phases. The governing equations are solved numerically by the SPH method. The phase change between the liquid and gas phases are simulated directly without using any phase change models. The numerical method is validated by comparing numerical results with analytical solutions for the vaporization of n-heptane drops at different temperatures. Using the SPH method, the processes of n-heptane drops impacting on a solid wall with different temperatures are studied numerically. The results show that the size of the film formed by drop impact decreases when temperature increases. When the temperature is high enough, the drop will rebound.

## 1. Introduction

Drop-wall interaction is of importance in combustion engines. The impact of a drop generally goes through three different phases, namely kinematic, spreading, relaxation or wetting [1, 2]. During the kinematic phase, the rate of spread depends on impact velocity and drop diameter. As seen from experimental results [3], in the first 1 ms, the spread factor is independent of temperature. This is followed by the spreading phase, when viscosity and surface tension slow down the liquid motion. This is followed by either the receding or wetting phase, depending on the wettability or the equilibrium contact angle at the liquid-solid-gas interface. Harlow and Shannon [4] created the Marker-and-Cell (MAC) finite difference method, which was among the first numerical model developed for studying droplet impact. It precludes the effects of surface tension and viscosity and therefore has limited utility in predicting the kinematic phase only. Experiments illustrated the same phenomena through a quantitative comparison [3]. Models including the capillary effects were later introduced [5, 6]. Simulating the no-slip condition at the solid-liquid interface led to force singularity [5] which was avoided by using a slip model [6]. A slip model is applied to the simulations reported in this paper, which involves using a very high value of liquid-wall viscosity (10-20 times more than liquid viscosity) such that the high frictional force at the interface replicates the no-slip condition to a satisfactory degree.

Sub Topic: Heterogeneous Combustion, Sprays & Droplets

## 2. SPH Methods

The smoothed particle hydrodynamics (SPH) method is applied to simulate the process of drop-wall interactions and drop vaporization. SPH is a meshless particle method [7, 8]. In SPH, the value of a function $f$ at a position $r$ can be calculated by the following particle summation.

$$f(r) = \sum_{j=1}^{N} \frac{m_j}{\rho_j} f(r_j) W(r - r_j, h) \tag{1}$$

Here $m$ and $\rho$ are the mass and density of a particle, respectively. $W$ is a kernel function and $h$ is a smoothing length to control the size of the summation domain. In this work, the following kernel function is used for stability benefits [9, 10].

$$W(r_a - r_b, h) \equiv W_{ab} = \alpha_d \begin{cases} s^3 - 6s + 6, & 0 \le s < 1 \\ (2-s)^3, & 1 \le s < 2 \\ 0, & 2 \le s \end{cases} \tag{2}$$

Here $s = |r_a - r_b|/h$. The parameter $\alpha_d$ has the values of $1/(3\pi h^2)$ and $15/(62\pi h^3)$ in two- and three-dimensions, respectively. A variable smoothing length is used to ensure sufficient number of particles in calculation of field properties [11].

With particle summation, the density of an SPH particle can be calculated as

$$\rho_a = \sum_b m_b W_{ab} \tag{3}$$

The momentum and energy equations are used to control the fluid flow.

$$\frac{d\mathbf{u}}{dt} = \frac{1}{\rho} \nabla \cdot \mathbf{S} + \mathbf{g} \tag{4}$$

$$\frac{de}{dt} = \frac{1}{\rho} \mathbf{S} : \nabla \mathbf{u} - \frac{1}{\rho} \nabla \cdot \mathbf{q} \tag{5}$$

Here $\mathbf{u}$ is the fluid velocity, $\mathbf{S} = -p\mathbf{I} + \sigma$ is tress tensor, $\mathbf{g}$ is the gravitation, $e$ is the special energy, $\mathbf{q} = -\kappa \nabla T$ is the heat flux, and $T$ is the fluid temperature. In the SPH method, the above equations can be discretized into particle equations as follows

$$\frac{d\mathbf{u}_a}{dt} = \sum_b m_b \left( \frac{\mathbf{S}_a}{\rho_a^2} + \frac{\mathbf{S}_b}{\rho_b^2} \right) \cdot \nabla_a W_{ab} + \mathbf{g}_a \tag{6}$$

$$\frac{de_a}{dt} = \sum_b m_b \left( \frac{\mathbf{S}_a}{\rho_a^2} + \frac{\mathbf{S}_b}{\rho_b^2} \right) : (\mathbf{u}_a - \mathbf{u}_b) \nabla_a W_{ab} - \sum_b m_b \left( \frac{\mathbf{q}_a}{\rho_a^2} + \frac{\mathbf{q}_b}{\rho_b^2} \right) \cdot \nabla_a W_{ab} \tag{7}$$

$$\mathbf{q}_a = -\kappa \sum_b 2 m_b \frac{T_b - T_a}{\rho_a + \rho_b} \nabla_a W_{ab} \tag{8}$$

The pressure is calculated using the Peng-Robinson equation of state.

$$p = f\left( \frac{RT}{V-b} - \frac{a}{V^2 + 2bV - b^2} \right) \tag{9}$$

Here $a = a_c[1 + k(1 - \sqrt{T_r})]^2$ and $T_r = T/T_c$. The temperature is determined from the following calorific equation.

$$T = (e + a\rho)/c \tag{10}$$

The value of $c$ is chosen such that it could predict the saturated liquid and vapor enthalpies accurately.

Sub Topic: Heterogeneous Combustion, Sprays & Droplets

Boundary condition at the wall is imposed by means of a wall force as follows.

$$\frac{du_a^w}{dt} = \begin{cases} 0, & y_a > y_r \\ -ka(y_a - y_w), & y_w < y_a \leq y_r \\ -kc(y_a - y_w), & y_a \leq y_w \end{cases} \quad (11)$$

This force acts on the liquid particles normal to the wall surface. The magnitude of this repulsion increases proportionally with depth (constant of proportionality = *kc*). An attractive force proportional to the height of the particle above the wall (constant of proportionality = *ka*), acts to pull the liquid particles towards the wall.

### 3. Results and Discussion

Fig. 1 compares the numerical results of saturated temperature and the analytical saturation line obtained by using Maxwell's equal-area rule and the Peng-Robinson equation of state. The liquid-vapor equilibrium state at different temperatures could be predicted with sufficient accuracy.

Fig. 2 illustrates the distribution of the SPH particles and the spatial variation of density. It can be seen that as temperature increases, more vapor is produced. It can also be seen that the liquid drop has a clear surface at low temperature ($T = 297$ K), because there is few vapor. As temperature increases, the interface between the liquid and gas phases become unclear. As temperature approached the critical temperature ($T = 545$ K), there is no clear distinction between the liquid and gas phases.

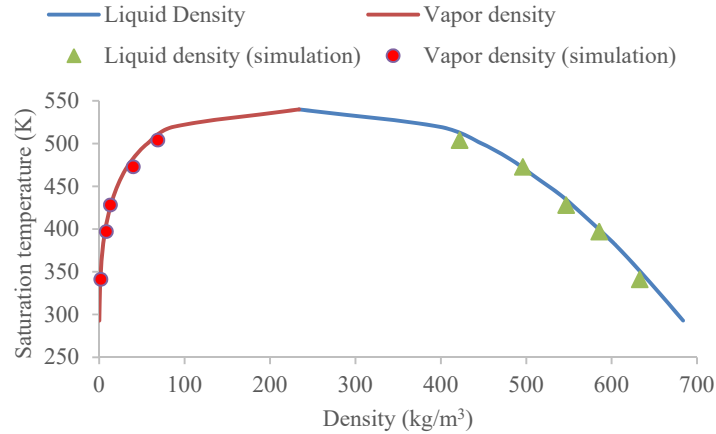

Figure 1: Comparison of saturated liquid and vapor densities of n-heptane obtained from experiments and SPH simulations

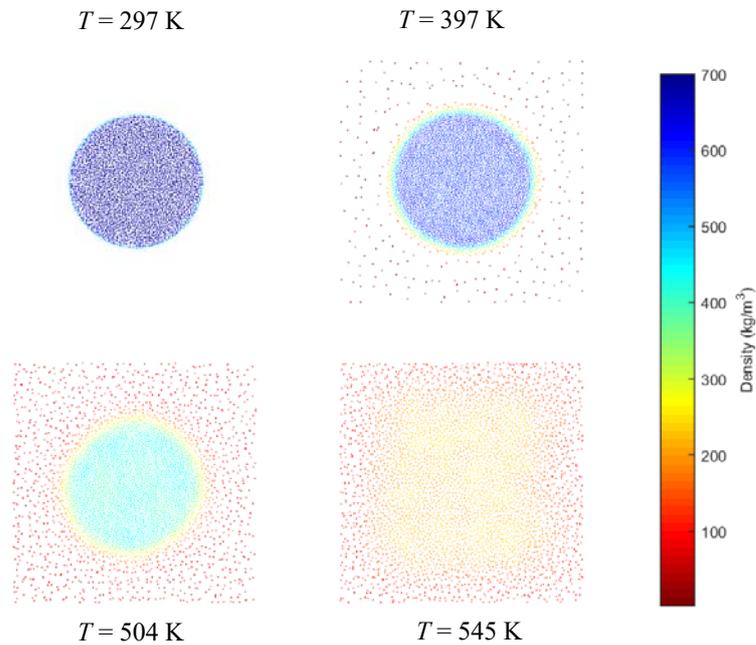

Figure 2: Particle and density distribution at different temperatures (n-heptane)

Fig. 3 shows the 2D simulation results of a 1.5 mm n-heptane drop impacting on a hot surface at different temperatures. It can be seen that the drop spreads on the wall and forms a film at room temperature. As the temperature increases, the size of the film decreases and the film thickness at the rim increases. However, the thickness at the center of the film decreases. The film even breaks when the temperature is high enough (Fig. 3, $T$ = 150 °C and 200 °C and $t$ = 6 ms). At the same time, more vapor is produced as the temperature increases.

Although the 2D simulation can predict the size decrease of the film with temperature, it cannot accurately capture the 3D effect of drop-wall interactions. For example, the rim of the film is a circle in a 3D case, and the surface tension in a 3D case is stronger than that in a 2D case. Fig. 4 shows the 3D simulation results of a 1.5 mm n-heptane drop impacting on a hot surface at different temperatures. Due to surface tension, the rim tends to shrink and to decrease its size. It can be seen in Fig. 4 that as the temperature increases, the thickness at the center of the film decreases but does not break. For the case with $T$ = 200 °C, the drop rebounds at the end.

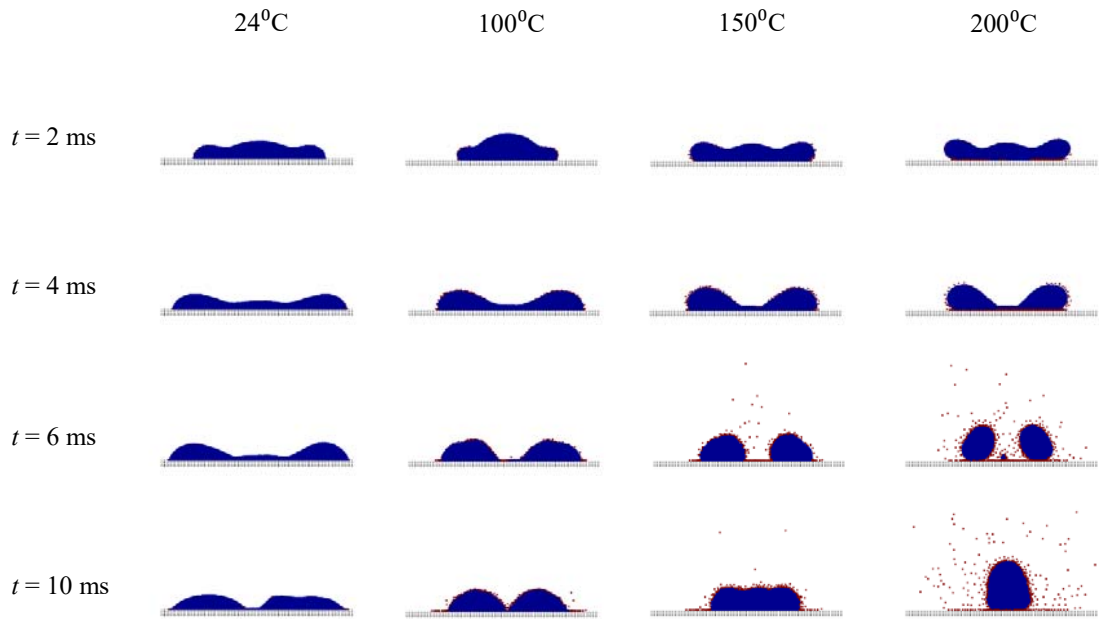

Figure 3: 2D simulation of impact and vaporization of a 1.5 mm n-heptane liquid drop at different wall temperatures (blue: liquid particles, red: vapor particles)

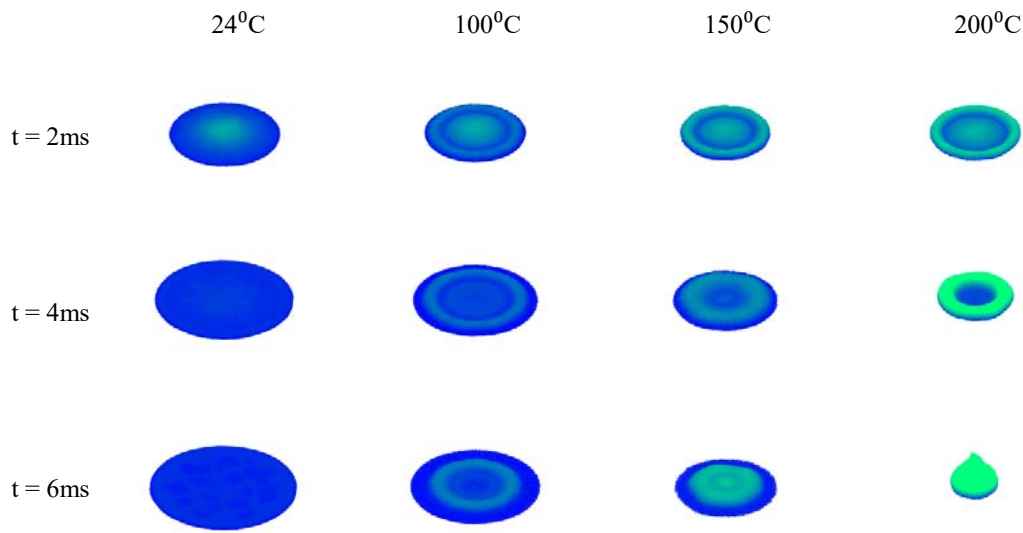

Figure 4: Isometric view of impact and vaporization of a 1.5 mm n-heptane liquid drop at different wall temperatures (vapor particles have been removed from the illustration to emphasize the liquid surface structure)

## 4. Conclusions

This paper presents an SPH method for simulating drop vaporization and impact on a hot surface. The conservation equations of momentum and energy are used to describe the fluid flow and heat transfer for both the liquid and gas phases. The Peng-Robinson equation of state are applied to predict the fluid pressure. The governing equations are solved numerically by the SPH method. The phase change between the liquid and gas phases are simulated directly without using any phase change models. In order to validate the present SPH method, the vaporization process of n-heptane drops is simulated. The numerical results agree well with the analytical solutions. The present numerical method is able to predict a higher vaporization rate as temperature increases. The processes of n-heptane drops impacting on a solid wall with different temperatures are simulated in two- and three-dimensions. The simulation predicts the spread, breakup, and rebound of the drop. The results also show the decrease in film size as temperature increases.

## 5. References

[1] M. Passandideh Fard, Y. Qiao, S. Chandra, J. Mostaghimi, The effect of surface tension and contact angle on the spreading of a droplet impacting on a substrate, ASME Fluids Engineering Conference (1995) 53-62.
[2] R. Rioboo, M. Marengo, C. Tropea, Time evolution of liquid drop impact onto solid, dry surfaces, Exp. Fluids 33(2002) 112-124.
[3] S. Chandra, C. Avedisian, On the collision of a droplet with a solid surface, Proceedings of the Royal Society of London A: Mathematical, Physical and Engineering Sciences (1991) 13-41.
[4] F.H. Harlow, J.P. Shannon, The splash of a liquid drop, Journal of Applied Physics 38(1967) 3855-3866.
[5] L. Hocking, A. Rivers, The spreading of a drop by capillary action, J. Fluid Mech. 121(1982) 425-442.
[6] E. Ramé, S. Garoff, On identifying the appropriate boundary conditions at a moving contact line: an experimental investigation, J. Fluid Mech. 230(1991) 97-116.
[7] L.B. Lucy, A numerical approach to the testing of the fission hypothesis, Astron. J. 82(1977) 1013-1024.
[8] R.A. Gingold, J.J. Monaghan, Smoothed particle hydrodynamics: theory and application to non-spherical stars, Mon. Not. R. Astron. Soc. 181(1977) 375-389.
[9] X. Yang, M. Liu, S. Peng, Smoothed particle hydrodynamics modeling of viscous liquid drop without tensile instability, Comput. Fluids 92(2014) 199-208.
[10] X.F. Yang, M.B. Liu, Improvement on stress instability in smoothed particle hydrodynamics, Acta Phys. Sin. 61(2012).
[11] J. Lattanzio, J. Monaghan, H. Pongracic, M. Schwarz, Interstellar cloud collisions, Mon. Not. R. Astron. Soc. 215(1985) 125-147.